\documentclass[journal=jpcc,manuscript=article]{achemso}

\usepackage[version=3]{mhchem} 
\usepackage[T1]{fontenc}       
\usepackage{pstricks,epsfig}
\usepackage{caption,float,natbib,setspace,xkeyval}

\title{Hybrid density-functional theory calculations of electronic and optical properties of mercaptocarboxylic acids on ZnO   $(10{\overline 1}0)$ surfaces}

\author{Dennis Franke}
\affiliation{Bremen Center for Computational Materials Science, University of Bremen, Am Fallturm 1, 28359 Bremen, Germany}
\email{defra@uni-bremen.de}
\author{Michael Lorke}
\email{mlorke@uni-bremen.de}
\author{Th.Frauenheim}
\affiliation{BCCMS, Universit\"at Bremen, Am Fallturm 1, 28359, Bremen, Germany}
\email{frauenheim@bccms.uni-bremen.de}
\author{A. L. Rosa}
\affiliation{Federal University of Goi\'as, Institute of Physics, Campus Samambaia, 74690-900, Goi\^ania, Goi\'as, Brazil}
\alsoaffiliation{BCCMS, Universit\"at Bremen, Am Fallturm 1, 28359, Bremen, Germany}
\email{andreialuisa@ufg.br}

\begin{document}

\begin{abstract}
  In this work we investigate the electronic properties of
  mercaptocarboxylic acids with several carbon chain lengths adsorbed
  on ZnO-(10-10) surfaces via density functional theory calculations
  using semi-local and hybrid exchange-correlation
  functionals. Amongst the investigated structures, we identify the
  monodentate adsorption mode to be stable. Moreover, this mode
  introduces optically active states in the ZnO gap, is further
  confirmed by the calculation of the dielectric function at PBE0 and
  TD-PBE0 levels. One interesting finding is that adsorption mode and
  the dielectric properties of the hybrid system are both rather
  insensitive to the chain length, since the acceptor molecular state
  is very localized on the sulphur atom. This indicates that even
  small molecules can be used to stabilize ZnO surface and to enhance
  its functionality for opto-electronic applications.
\end{abstract}

\maketitle

\section{Introduction}

Hybrid nanostructures made of organic and inorganic materials are of
great interest for applications in electronic and optoelectronics
devices. ZnO is a semiconductor and a low-cost
material with a wide band gap of 3.3 eV, high electron mobility, and
high absorbance in the UV range\,\cite{Materials:ZnO}. Besides, ZnO can be synthesized in several nanostructure shapes. ZnO nanowires are of
particular interest for electronic transport in
one-dimension\,\cite{ZLWang_Science07}. Because of the large
surface-to-volume ratio of ZnO nanowires, their optical and electrical
properties can be efficiently tailored by surface functionalization.
On the other hand, zero dimensional semiconductor nanostructures such
as colloidal quantum dots exhibit exceptional optical and
electronic properties and photo-stability. These nanostructures have a
size-dependent band gap which allows efficient absorption across the broad
solar spectrum\,\cite{NatComm2017,NatPhot2012,Semonin}. Recently
it has been proposed that surface functionalization of ZnO nanowires
using ligand-capped quantum dots lead to hybrid solar cells with high
absorption in a wide spectral
range\,\cite{Leschkies2007,Bley2015}. In such hybrid systems, functional
groups are attached on both quantum dot and nanowire surfaces.

Ab initio density-functional theory calculations of several small
functional groups on ZnO $(10{\overline 1}0)$ surfaces have been
pursued\,\cite{LeBahers_Langmuir,Kiss,Calzolari2,APL_Ney,PCCP,SurfSci}.
In our previous work, we have shown that -SH groups
form strong covalent bonds on ZnO $(10{\overline 1}0)$ surfaces with a
monodentate adsorption mode\,\cite{PCCP,JAP14}. Indeed thiols
have been investigated experimentally and found to bind strongly to ZnO non-polar surfaces. In particular, molecules with long ${\rm CH_2}$
chains are observed to be stable\,\cite{Hamers2011,Hamers2012}.

In this paper we have
investigated mercaptocarboxylic acids (MPA)-derived molecules ${\rm
  SH-(CH_2)_n-COOH}$, (n = 1,2,3,7) on ZnO  $(10{\overline 1}0)$
surfaces. Monodentate adsorption mode is found to be stable and
introduces intragap states which are responsible for changes in the
electronic and optical properties of ZnO. Furthermore, adsorption modes and
optical properties of the hybrid interfaces are not sensitive to the
carbon chain length, which suggests that even small molecules can be used to
stabilize and modify the ZnO surface properties.

\section{Computational Details}

The calculations were performed using density functional theory (DFT)
as implemented in the Vienna ab initio simulation package
(VASP)\,\cite{VASP:3,VASP:4}. The projected augmented wave method has
been used \cite{Kresse:99,Bloechl}. The atomic structures were
optimized using the PBE functional\,\cite{Perdew:96}. To get better
description of the electronic structure, the PBE0 hybrid
functional\,\cite{PBE0} has been employed. This functional can
reproduce reasonably well the ZnO band gap\,\cite{CRC}. A plane wave basis with an
energy cutoff of ${\rm E_{cut}=400\,eV}$ and a $(1\times4\times4)$ Monkhorst-Pack {\bf k}-point sampling has been used.  The dielectric function has been
calculated in the independent particle approximation with the PBE0 functional
(IPA-PBE0). Additionally time-dependent density-functional theory with the PBE0
functional (TD-PBE0) has been used to include excitonic effects. In this case, only the $\Gamma$ point has been considered. The surfaces have been modeled using the slab approach with a $(3\times2)$ surface supercell containing 96 atoms (48 ZnO units) units and a single MPA molecule. 

\section{Results}

\begin{figure*}[ht!]
\centering
\includegraphics[width=0.7\textwidth,clip]{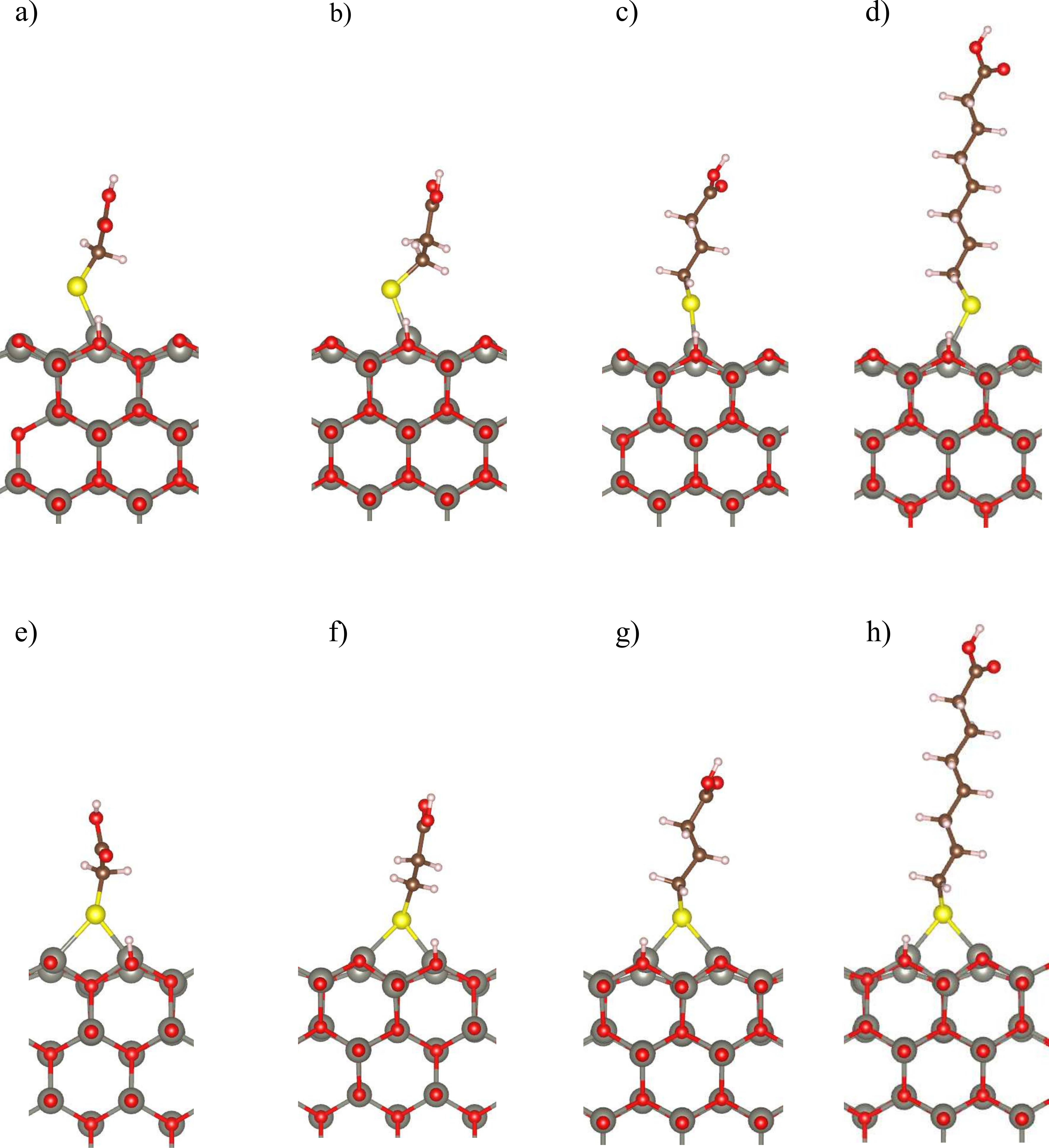}
\caption{Optimized structures with PBE for the ($10\bar{1}0$) ZnO surface functionalized 
with MPA with a monodentate binding mode a) SH-(CH$_{2}$)$_{n}$, a) n= 1, b) n=2, c) n=3, d) n=7 and bidentate mode e) n= 1, f) n=2, g) n=3 and h) n=7.}
\label{Geo}
\end{figure*}

We have explored several geometries for the MPA molecules on the ZnO
($10\bar{1}0$) surfaces. The ground-state geometries for
SH-(CH$_2)_n$-COOH on ZnO are shown in Fig.~\ref{Geo} (a)-(d) for both
monodentate and Fig.~\ref{Geo} (e)-(h) bidentate binding modes. These
geometries were optimized at the PBE level, as explained in our
previous calculations\,\cite{PCCP,JAP14}. The preferred adsorption
mode is the monodentate mode with the molecule sitting at top
sites. This mode is more stable against bridge and hollow
positions. In both cases the MPA molecules adsorb in a dissociative
manner.  Table\,\ref{table_1010} shows the bond lengths between the
sulphur atom of the molecule and the outermost zinc atom of the ZnO
surface $d_{\rm Zn-S}$({\AA}),the relaxation of the zinc atom which is
bonded to the sulphur atom relative to the zinc atoms of the ZnO
surface $\Delta z$, the energy difference $\Delta E$ between the
monodentate and bidentate configurations and the adsorption energy of
the MPA molecules ${\rm E_{ads}}$(eV) on the ZnO surface. The
monodentate mode is the most stable configuration for all investigated
carbon chain lengths, diminishing as the chain length increases. The
molecules sit relatively upright on the surface. The Zn-S bond length
and the buckling remains almost constant for all the structures,
indicating that the interaction of the molecules with the surface is
strong (around 1\,eV), as it can be seen from the adsorption energies in Table
\ref{table_1010}. In this work we have consider a low
coverage regime. In our case we have investigated a coverage of 1
molecule/nm$^2$, but densities up to 4.6 molecules/nm$^2$ have been
reported\cite{Hamers2012}. Larger coverage values need to be investigated in
order to see whether the monodentate mode remains stable against the
bidentate mode.

\begin{table}[ht!]
\caption{\label{table_1010} Outward relaxation of the surface zinc atom ${\rm \Delta z}$, Zn-S bond length $d_{\rm Zn-S}$, energy difference ${\rm \Delta E}$ between the monodentate and bidentate modes and adsorption energy  for the monodentate mode ${\rm E_{ads}=E_{tot}^{MPA/ZnO}-E_{tot}^{bare~surf}-E_{tot}^{MPA}}$ of 
SH-(CH$_{2}$)$_{n}$-COOH (n=1,2,3,7) molecules on ZnO  ($10\bar{1}0$) surfaces. ${\rm E_{tot}^{MPA/ZnO}}$, ${\rm E_{tot}^{bare~surf}}$ and ${\rm E_{tot}^{MPA}}$ are the total energy of the hybrid MPA/ZnO interface, of the bare ZnO  ($10\bar{1}0$) surface and of the MPA-derived molecules with different chain lengths, respectively.}
\begin{center}
\begin{tabular*}{1\textwidth}{@{\extracolsep{\fill}}lcccc}
\hline
chain length $n$           &   ${\rm \Delta z}$({\AA}) & $d_{\rm Zn-S}$({\AA})  &  ${\rm \Delta E}$(eV) & ${\rm E_{ads}}$(eV) \\
\hline
1           &   0.28   &   2.26   & -2.20  & 1.00 \\
2           &   0.35   &   2.23  &-1.90  & 1.00\\
3           &   0.28   & 2.26   & -1.10  & 1.10\\
7           &   0.28   & 2.26   &   -0.70  & 1.10 \\
\hline
\end{tabular*}
\end{center}
\end{table}

\begin{figure}[h]
\includegraphics[width=0.7\columnwidth,clip]{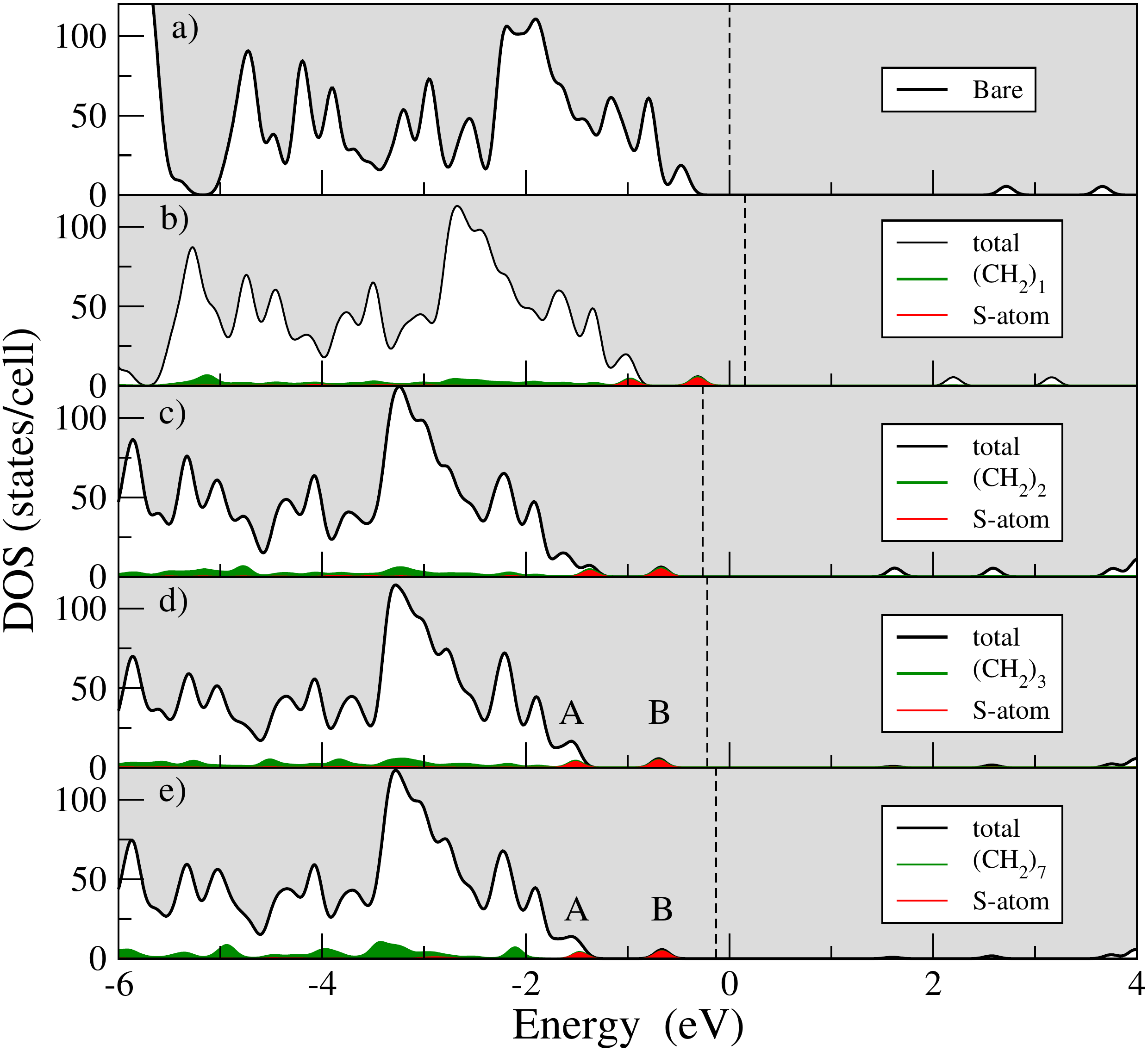}\\
\includegraphics[width=0.35\columnwidth,height=4cm,keepaspectratio,clip=]{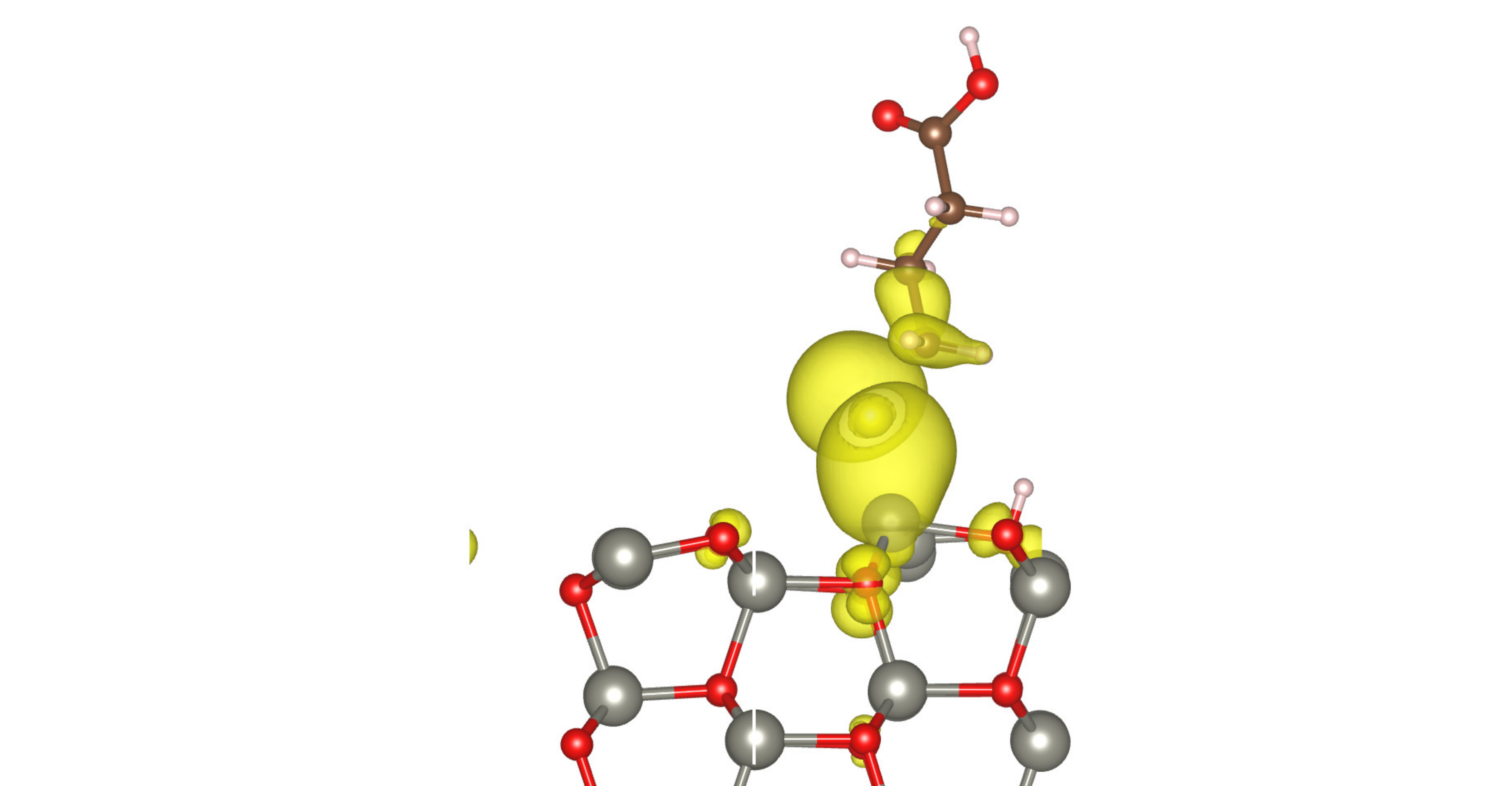}
\includegraphics[width=0.35\columnwidth,height=4cm,keepaspectratio,clip=]{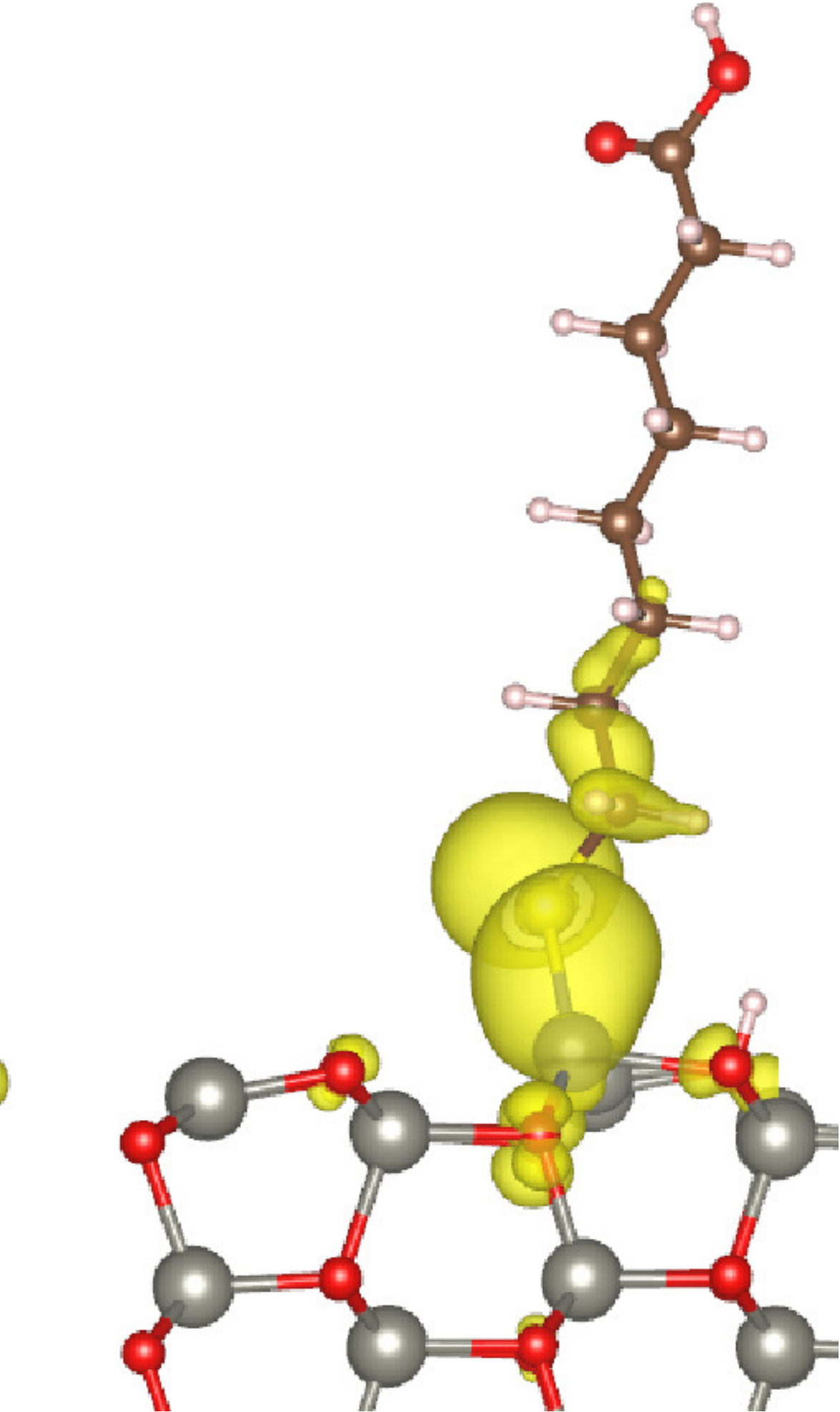}\\
\includegraphics[width=0.35\columnwidth,height=4cm,keepaspectratio,clip=]{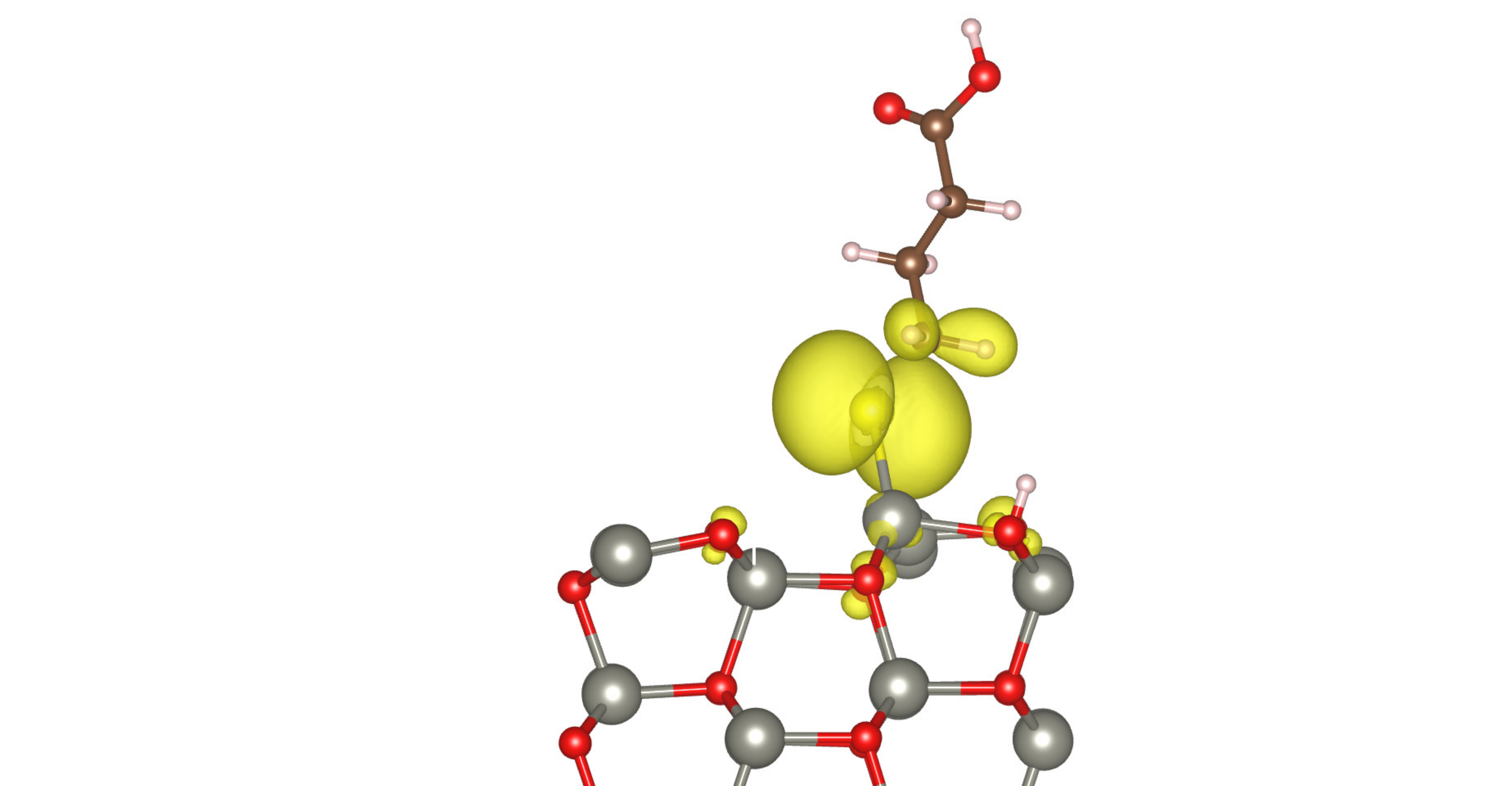}
\includegraphics[width=0.35\columnwidth,height=4cm,keepaspectratio,clip=]{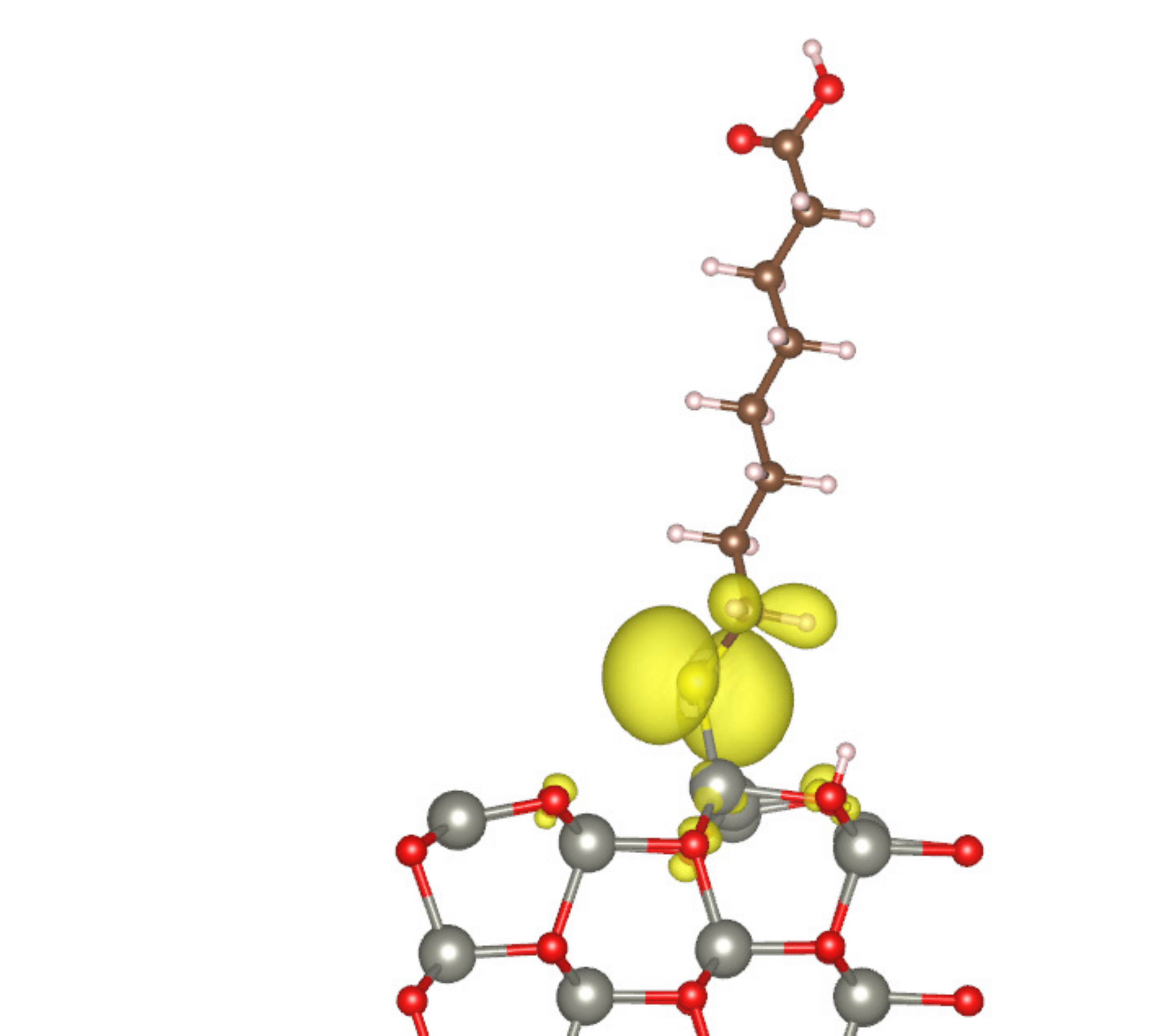}
\caption{(a)-(e) Total and projected DOS for the bare and modified
  surfaces with MPA in a monodentate binding mode.  The black and
  green lines represent the total DOS and its projection onto
  molecular states, respectively. The Fermi energy is denoted by
  dashed lines. The figures below the DOS show the band decomposed
  charge density at the $\Gamma$ point. The figures on the left correspond to the highest-minus-one-occupied orbital in SH-(CH$_{2}$)$_{3}$-COOH and   SH-(CH$_{2}$)$_{7}$-COOH on the ZnO surface. The figures on the right correspond to the highest occupied orbital in  SH-(CH$_{2}$)$_{3}$-COOH and SH-(CH$_{2}$)$_{7}$-COOH on
  ($10\bar{1}0$) ZnO surface. The A and B states are indicated in the DOS (c) and (e).}
\label{fig:electronic}
\end{figure}

The total and projected density-of-states (PDOS) calculated at the
PBE0 level are shown in Fig.~\ref{fig:electronic}. The eigenvalues
have been aligned according to the electrostatic potential of the
vacuum region as described in Ref.~\cite{Pasquarello:09}. The zero of
the energy is set at the highest occupied state of the bare ZnO
($10\bar{1}0$) surface shown in Fig.~\ref{fig:electronic}(a) as
explained in our previous investigations Ref.~\cite{PCCP,JAP14}. For
all functionalized structures, intra-gap states are found, as it can
be seen in Figs.~\ref{fig:electronic}(b)-(e). These states are mainly
localized on the sulphur atom and at the first -CH$_2$ group of the
molecule. The difference between the state located at valence band
maximum (VBM) and the occupied state inside the gap remains unchanged,
confirming that the electronic structure of the hybrid system is
dominated by the interaction between atoms close to the surface. In
order to provide further information on charge localization, we have
calculated the band decomposed partial charge density at the $\Gamma$
point, as shown in Fig.\,\ref{fig:electronic} for ${\rm
  SH-(CH_2)_3-COOH}$ and ${\rm SH-(CH_2)_7-COOH}$ molecules.  The
figures on the left correspond to the highest-minus-one-occupied
orbital in SH-(CH$_{2}$)$_{3}$-COOH (top left) and
SH-(CH$_{2}$)$_{7}$-COOH (top right) on the ($10\bar{1}0$) ZnO
surface.  The figures on the right correspond to the highest occupied
orbital in SH-(CH$_{2}$)$_{3}$-COOH (top left) and
SH-(CH$_{2}$)$_{7}$-COOH (top right) on the ($10\bar{1}0$) ZnO
surface. The corresponding A and B states are indicated in the
Figs.~\ref{fig:electronic} (c) and (e).  Both molecular states (PDOS
  shown in green) are clearly dominated by contributions from
  sulphur-$p$ orbitals from the molecules (PDOS shown in red), with
  small hybridization both with the ZnO surface and the neighboring
  CH$_2$ groups. This feature is found for all investigated
  structures. The band decomposed partial charge for the MPA
  molecules with smaller chains (n=1 and n=2) show a similar
  behavior (see Fig. S1 in Supporting Information). This indicates that MPA
  molecules with -SH groups attached to the ZnO surface, the -CH$_2$ groups
  plays a minor role in the charge transfer process and consequently
  on the electronic structure of the system.

\begin{figure}[h]
\centering
\includegraphics[width=1\columnwidth,clip]{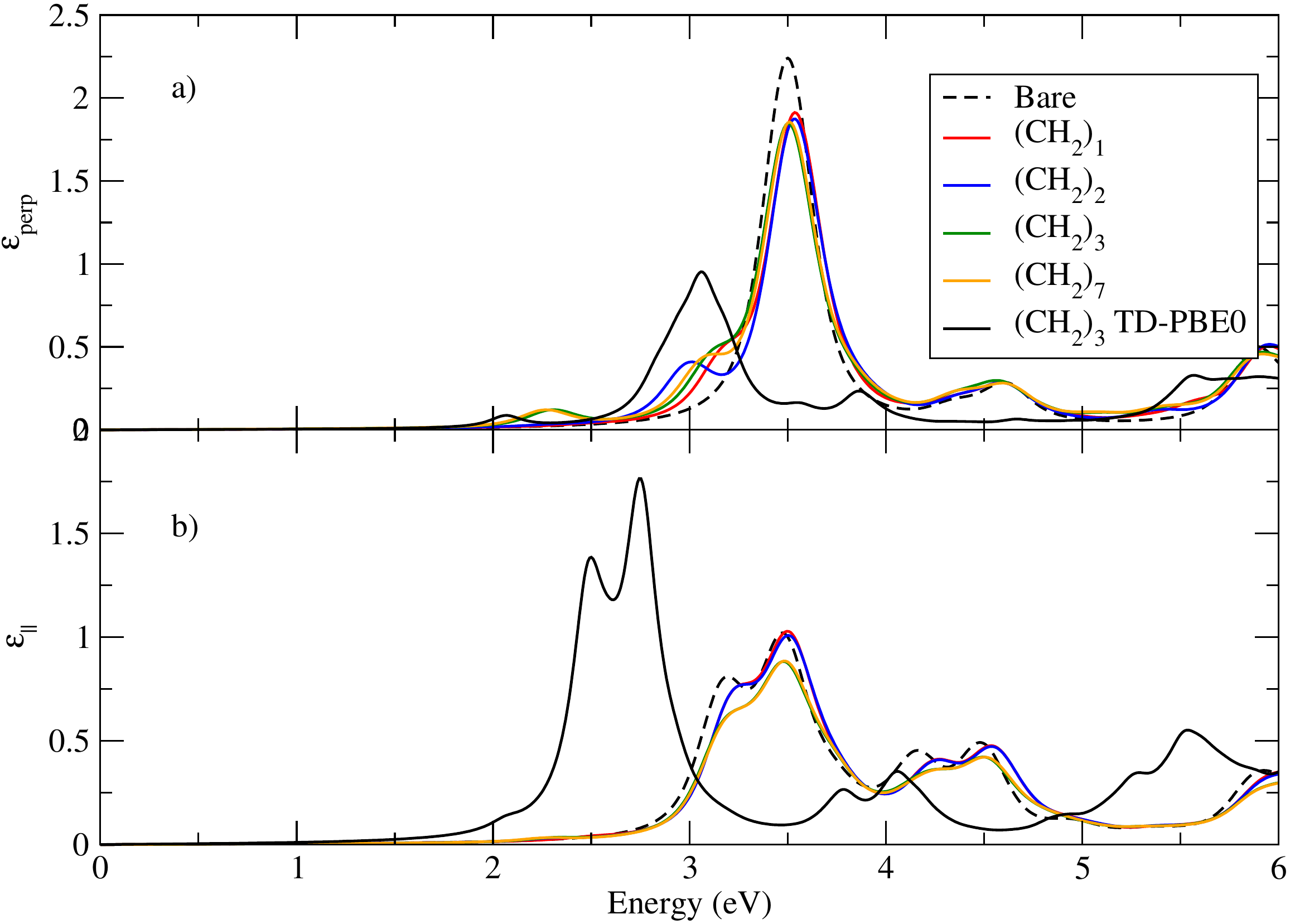}
\caption{Dielectric function for the bare and modified surfaces with a monodentate binding mode, 
shown are $\varepsilon_{\perp}$ (a) and $\varepsilon_{||}$ (b).}
\label{fig:diel}
\end{figure}

The presence of the intra-gap states raises the question whether these
systems can be used to enhance or modify the optoelectronic properties
of the ZnO non-polar surfaces. Therefore we have calculated the
dielectric function of the bare and functionalized surfaces in the
IPA\,\cite{Gadjos2006}.  The $\varepsilon_{\perp}$ component shown in
Fig.~\ref{fig:diel} (a) corresponds to the propagation of the
electromagnetic field perpendicular to the surface while
$\varepsilon_{||}$ in Fig.~\ref{fig:diel} (b) corresponds to the field
being parallel to the surface.  We find that in both cases the
intra-gap states are optically active. Especially for
$\varepsilon_{\perp}$ a peak around 2.3\,eV is clearly visible for
larger chains, stemming from transitions between the molecular state B
and the ZnO conduction band.  Additionally, the spectra reveal a
shoulder at the low energetic side of the band-to-band transition,
that corresponds to a transition between the state A and the ZnO
conduction band. For smaller carbon chain lengths such as for n=2,
only the latter is found. This is because the charge is less localized
across the surface and extends to the functional group -COOH located
at the upper end of the molecule. Therefore, depending on the intended
charge localization one can choose the proper carbon chain length to
be used. 

The discussion above does not include the presence of excitonic
effects. However, more accurate predictions of materials for
optoelectronic applications requires the inclusion of the
electron-hole pair, which can be achieved using theoretical methods at
different approximations, such as GW + Bethe-Salpether equation (BSE)
or TD-DFT\,\cite{Onida}. Previous theoretical investigations reported
excitation energies for catechol and dopamine interactions with
TiO$_2$ surfaces using TD-DFT\,\cite{Luppi2016}. Therefore, in
addition to the PBE0 calculations for the dielectric function, we have
performed TD-DFT calculations\,\cite{Onida} so that occupied and empty
one-electron states are calculated using the PBE0 functional. As
discussed in Ref.\,\cite{Onida}, this approach should lead to
excitation spectra which are improved compared to the IPA-PBE0
approach and equivalent to the solution of the BSE equation. In this
manner, excitonic effects can be captured to a certain extent. The
dielectric function is then obtained by solving the Casida's
equation\,\cite{Casida}. Our results for the MPA with n = 3 are shown
in Fig.~\ref{fig:diel} (a) for $\varepsilon_{\perp}$ and (b) for
$\varepsilon_{\parallel}$.  Our results show that the inclusion of
many-body effects in TD-PBE0 causes a red shift of the
spectrum. Further experiments to validate our results are
needed. Also, it would be useful to have the absorption spectrum
calculated with GW+BSE, but this is still prohibitive due to the high
computational costs. Furthermore, we would like to point out that,
even though the spectra are calculated here in the IPA, the inclusion
of excitonic (ladder) effects is not expected to change the physical
picture significantly, as the exciton binding energies between
localized molecular states and delocalized conduction band states
should be weak due to a small wave function overlap in the
corresponding exchange Coulomb matrix elements.  Hence, we believe
that the absorption due to the presence of -SH groups states around
2.3\,eV can be used either to enhance the efficiency of solar cells
based on functionalized semiconductor surfaces. Usually, the nanowire
growth direction is the [0001] with the non-polar surfaces exposed.
Therefore, the field propagation would be most beneficial if it
naturally happens along the nanowires surfaces.

\section{Conclusions}

We have investigated adsorption of MPA-derived molecules with several
carbon chain lengths using semi-local and hybrid functionals.  We have
identified the monodentate adsorption mode as the most stable one and
have further shown that electronic and optical properties of these
hybrid system are not very sensitive to the chain length. The acceptor
state located in the band gap is very localized on the molecular
orbitals, mainly on the sulphur atom. This indicates that even small
molecules can be used to stabilize ZnO non-polar surfaces and tune their
dielectric properties. Our results corroborate well with experimental
findings which have suggested that this system is promising for
optoelectronic applications.

\begin{acknowledgement}

The authors acknowledge funding by the DFG research group FOR 1616 ``Dynamics and Interactions of Semiconductor Nanowires for Optoelectronics'', CNPq and FAPEG.

\end{acknowledgement}

\begin{suppinfo}

\end{suppinfo}

\bibliography{references}

\providecommand{\latin}[1]{#1}
\providecommand*\mcitethebibliography{\thebibliography}
\csname @ifundefined\endcsname{endmcitethebibliography}
  {\let\endmcitethebibliography\endthebibliography}{}
\begin{mcitethebibliography}{29}
\providecommand*\natexlab[1]{#1}
\providecommand*\mciteSetBstSublistMode[1]{}
\providecommand*\mciteSetBstMaxWidthForm[2]{}
\providecommand*\mciteBstWouldAddEndPuncttrue
  {\def\EndOfBibitem{\unskip.}}
\providecommand*\mciteBstWouldAddEndPunctfalse
  {\let\EndOfBibitem\relax}
\providecommand*\mciteSetBstMidEndSepPunct[3]{}
\providecommand*\mciteSetBstSublistLabelBeginEnd[3]{}
\providecommand*\EndOfBibitem{}
\mciteSetBstSublistMode{f}
\mciteSetBstMaxWidthForm{subitem}{(\alph{mcitesubitemcount})}
\mciteSetBstSublistLabelBeginEnd
  {\mcitemaxwidthsubitemform\space}
  {\relax}
  {\relax}

\bibitem[Kolodziejczak-Radzimska and Jesionowski(2014)Kolodziejczak-Radzimska,
  and Jesionowski]{Materials:ZnO}
Kolodziejczak-Radzimska,~A.; Jesionowski,~T. \emph{Materials} \textbf{2014},
  \emph{7}, 2833\relax
\mciteBstWouldAddEndPuncttrue
\mciteSetBstMidEndSepPunct{\mcitedefaultmidpunct}
{\mcitedefaultendpunct}{\mcitedefaultseppunct}\relax
\EndOfBibitem
\bibitem[Wang and Song(2006)Wang, and Song]{ZLWang_Science07}
Wang,~Z.; Song,~J. \emph{Science} \textbf{2006}, \emph{312}, 242\relax
\mciteBstWouldAddEndPuncttrue
\mciteSetBstMidEndSepPunct{\mcitedefaultmidpunct}
{\mcitedefaultendpunct}{\mcitedefaultseppunct}\relax
\EndOfBibitem
\bibitem[Yang \latin{et~al.}(2017)Yang, Fan, Proppe, de~Arquer, Rossouw,
  Voznyy, X, Liu, Walters, Quintero-Bermudez, Sun, Hoogland, and. S.~O.~Kelley,
  and Sargent]{NatComm2017}
Yang,~Z.; Fan,~J.~Z.; Proppe,~A.~H.; de~Arquer,~F. P.~G.; Rossouw,~D.;
  Voznyy,~O.; X,~L.; Liu,~M.; Walters,~G.; Quintero-Bermudez,~R.; Sun,~B.;
  Hoogland,~S.; and. S.~O.~Kelley,~G. A.~B.; Sargent,~E.~H. \emph{Nat. Comm.}
  \textbf{2017}, \emph{8}, 1325\relax
\mciteBstWouldAddEndPuncttrue
\mciteSetBstMidEndSepPunct{\mcitedefaultmidpunct}
{\mcitedefaultendpunct}{\mcitedefaultseppunct}\relax
\EndOfBibitem
\bibitem[Sargent(2012)]{NatPhot2012}
Sargent,~E.~H. \emph{Nat. Photon.} \textbf{2012}, \emph{6}, 133\relax
\mciteBstWouldAddEndPuncttrue
\mciteSetBstMidEndSepPunct{\mcitedefaultmidpunct}
{\mcitedefaultendpunct}{\mcitedefaultseppunct}\relax
\EndOfBibitem
\bibitem[Semonin \latin{et~al.}(2012)Semonin, M.Luther, and C.Beard]{Semonin}
Semonin,~O.~E.; M.Luther,~J.; C.Beard,~M. \emph{Materials Today} \textbf{2012},
  \emph{15}, 508\relax
\mciteBstWouldAddEndPuncttrue
\mciteSetBstMidEndSepPunct{\mcitedefaultmidpunct}
{\mcitedefaultendpunct}{\mcitedefaultseppunct}\relax
\EndOfBibitem
\bibitem[Leschkies \latin{et~al.}(2007)Leschkies, Divakar, Basu, Enache-Pommer,
  Boercker, Carter, Kortshagen, Norris, and Aydil]{Leschkies2007}
Leschkies,~K.~S.; Divakar,~R.; Basu,~J.; Enache-Pommer,~E.; Boercker,~J.~E.;
  Carter,~C.~B.; Kortshagen,~U.~R.; Norris,~D.~J.; Aydil,~E.~S. \emph{Nano
  Lett.} \textbf{2007}, 1793\relax
\mciteBstWouldAddEndPuncttrue
\mciteSetBstMidEndSepPunct{\mcitedefaultmidpunct}
{\mcitedefaultendpunct}{\mcitedefaultseppunct}\relax
\EndOfBibitem
\bibitem[Bley \latin{et~al.}(2015)Bley, Diez, Albrecht, Resch, Waldvogel,
  Menzel, Zacharias, Gutowski, and Voss]{Bley2015}
Bley,~S.; Diez,~M.; Albrecht,~F.; Resch,~S.; Waldvogel,~S.~R.; Menzel,~A.;
  Zacharias,~M.; Gutowski,~J.; Voss,~T. \emph{J. Phys. Chem. C} \textbf{2015},
  \emph{119}, 15627\relax
\mciteBstWouldAddEndPuncttrue
\mciteSetBstMidEndSepPunct{\mcitedefaultmidpunct}
{\mcitedefaultendpunct}{\mcitedefaultseppunct}\relax
\EndOfBibitem
\bibitem[Bahers \latin{et~al.}(2011)Bahers, Pauport\'e, Labat, and
  Ciofini]{LeBahers_Langmuir}
Bahers,~T.~L.; Pauport\'e,~T.; Labat,~F.; Ciofini,~I. \emph{Langmuir}
  \textbf{2011}, \emph{27}, 3442\relax
\mciteBstWouldAddEndPuncttrue
\mciteSetBstMidEndSepPunct{\mcitedefaultmidpunct}
{\mcitedefaultendpunct}{\mcitedefaultseppunct}\relax
\EndOfBibitem
\bibitem[Kiss \latin{et~al.}(2011)Kiss, Langenberg, Silber, Traeger, Jin, Qiu,
  Wang, Meyer, and W\"oll]{Kiss}
Kiss,~J.; Langenberg,~D.; Silber,~D.; Traeger,~F.; Jin,~L.; Qiu,~H.; Wang,~Y.;
  Meyer,~B.; W\"oll,~C. \emph{J. Phys. Chem. A} \textbf{2011}, \emph{115},
  7180\relax
\mciteBstWouldAddEndPuncttrue
\mciteSetBstMidEndSepPunct{\mcitedefaultmidpunct}
{\mcitedefaultendpunct}{\mcitedefaultseppunct}\relax
\EndOfBibitem
\bibitem[Calzolari \latin{et~al.}(2011)Calzolari, Ruini, and
  Catellani]{Calzolari2}
Calzolari,~A.; Ruini,~A.; Catellani,~A. \emph{J. Am. Chem. Soc.} \textbf{2011},
  \emph{133}, 5893\relax
\mciteBstWouldAddEndPuncttrue
\mciteSetBstMidEndSepPunct{\mcitedefaultmidpunct}
{\mcitedefaultendpunct}{\mcitedefaultseppunct}\relax
\EndOfBibitem
\bibitem[Moreira \latin{et~al.}(2009)Moreira, Rosa, and Frauenheim]{APL_Ney}
Moreira,~N.~H.; Rosa,~A.~L.; Frauenheim,~T. \emph{Appl. Phys. Lett.}
  \textbf{2009}, \emph{94}, 193109\relax
\mciteBstWouldAddEndPuncttrue
\mciteSetBstMidEndSepPunct{\mcitedefaultmidpunct}
{\mcitedefaultendpunct}{\mcitedefaultseppunct}\relax
\EndOfBibitem
\bibitem[Moreira \latin{et~al.}(2012)Moreira, Dominguez, Frauenheim, and
  da~Rosa]{PCCP}
Moreira,~N.~H.; Dominguez,~A.; Frauenheim,~T.; da~Rosa,~A.~L. \emph{Phys. Chem.
  Chem. Phys.} \textbf{2012}, \emph{14}, 15445\relax
\mciteBstWouldAddEndPuncttrue
\mciteSetBstMidEndSepPunct{\mcitedefaultmidpunct}
{\mcitedefaultendpunct}{\mcitedefaultseppunct}\relax
\EndOfBibitem
\bibitem[Shi \latin{et~al.}(2012)Shi, Xu, Hove, Moreira, Rosa, and
  Frauenheim]{SurfSci}
Shi,~X.~Q.; Xu,~H.; Hove,~M.~V.; Moreira,~N.; Rosa,~A.; Frauenheim,~T.
  \emph{Surf. Sci.} \textbf{2012}, \emph{606}, 289\relax
\mciteBstWouldAddEndPuncttrue
\mciteSetBstMidEndSepPunct{\mcitedefaultmidpunct}
{\mcitedefaultendpunct}{\mcitedefaultseppunct}\relax
\EndOfBibitem
\bibitem[Dominguez \latin{et~al.}(2014)Dominguez, Lorke, Schoenhalz, Rosa,
  Frauenheim, Rocha, and Dalpian]{JAP14}
Dominguez,~A.; Lorke,~M.; Schoenhalz,~A.~L.; Rosa,~A.~L.; Frauenheim,~T.;
  Rocha,~A.~R.; Dalpian,~G.~M. \emph{J. Appl. Phys.} \textbf{2014}, \emph{115},
  203720\relax
\mciteBstWouldAddEndPuncttrue
\mciteSetBstMidEndSepPunct{\mcitedefaultmidpunct}
{\mcitedefaultendpunct}{\mcitedefaultseppunct}\relax
\EndOfBibitem
\bibitem[R.~E.~Ruther \latin{et~al.}(2011)R.~E.~Ruther, Huhn, Gomez-Zayas, and
  Hamers]{Hamers2011}
R.~E.~Ruther,~R.~F.; Huhn,~A.~M.; Gomez-Zayas,~J.; Hamers,~R.~J.
  \emph{Langmuir} \textbf{2011}, \emph{27}, 10604\relax
\mciteBstWouldAddEndPuncttrue
\mciteSetBstMidEndSepPunct{\mcitedefaultmidpunct}
{\mcitedefaultendpunct}{\mcitedefaultseppunct}\relax
\EndOfBibitem
\bibitem[Chen \latin{et~al.}(2012)Chen, Ruther, Tan, Bishop, and
  Hamers]{Hamers2012}
Chen,~J.; Ruther,~R.~E.; Tan,~Y.; Bishop,~L.~M.; Hamers,~R.~J. \emph{Langmuir}
  \textbf{2012}, \emph{28}, 10437\relax
\mciteBstWouldAddEndPuncttrue
\mciteSetBstMidEndSepPunct{\mcitedefaultmidpunct}
{\mcitedefaultendpunct}{\mcitedefaultseppunct}\relax
\EndOfBibitem
\bibitem[Kresse and Furthm\"uller(1996)Kresse, and Furthm\"uller]{VASP:3}
Kresse,~G.; Furthm\"uller,~J. \emph{Comput. Mat. Sci.} \textbf{1996}, \emph{6},
  15\relax
\mciteBstWouldAddEndPuncttrue
\mciteSetBstMidEndSepPunct{\mcitedefaultmidpunct}
{\mcitedefaultendpunct}{\mcitedefaultseppunct}\relax
\EndOfBibitem
\bibitem[Kresse and Furthm\"uller(1996)Kresse, and Furthm\"uller]{VASP:4}
Kresse,~G.; Furthm\"uller,~J. \emph{Phys. Rev. B} \textbf{1996}, \emph{54},
  11169\relax
\mciteBstWouldAddEndPuncttrue
\mciteSetBstMidEndSepPunct{\mcitedefaultmidpunct}
{\mcitedefaultendpunct}{\mcitedefaultseppunct}\relax
\EndOfBibitem
\bibitem[Kresse and Joubert(1999)Kresse, and Joubert]{Kresse:99}
Kresse,~G.; Joubert,~D. \emph{Phys. Rev. B} \textbf{1999}, \emph{59},
  1758\relax
\mciteBstWouldAddEndPuncttrue
\mciteSetBstMidEndSepPunct{\mcitedefaultmidpunct}
{\mcitedefaultendpunct}{\mcitedefaultseppunct}\relax
\EndOfBibitem
\bibitem[Bloechl(1994)]{Bloechl}
Bloechl,~P. \emph{Phys. Rev. B} \textbf{1994}, \emph{50}, 17953\relax
\mciteBstWouldAddEndPuncttrue
\mciteSetBstMidEndSepPunct{\mcitedefaultmidpunct}
{\mcitedefaultendpunct}{\mcitedefaultseppunct}\relax
\EndOfBibitem
\bibitem[Perdew \latin{et~al.}(1996)Perdew, Burke, and Ernzerhof]{Perdew:96}
Perdew,~J.~P.; Burke,~K.; Ernzerhof,~M. \emph{Phys. Rev. Lett} \textbf{1996},
  \emph{77}, 3865--3868\relax
\mciteBstWouldAddEndPuncttrue
\mciteSetBstMidEndSepPunct{\mcitedefaultmidpunct}
{\mcitedefaultendpunct}{\mcitedefaultseppunct}\relax
\EndOfBibitem
\bibitem[Adamo and Barone(1999)Adamo, and Barone]{PBE0}
Adamo,~C.; Barone,~V. \emph{J. Chem. Phys.} \textbf{1999}, \emph{110},
  6158\relax
\mciteBstWouldAddEndPuncttrue
\mciteSetBstMidEndSepPunct{\mcitedefaultmidpunct}
{\mcitedefaultendpunct}{\mcitedefaultseppunct}\relax
\EndOfBibitem
\bibitem[CRC(1977)]{CRC}
\emph{CRC Handbook of Chemistry and Physics}, 58th ed.; 1977\relax
\mciteBstWouldAddEndPuncttrue
\mciteSetBstMidEndSepPunct{\mcitedefaultmidpunct}
{\mcitedefaultendpunct}{\mcitedefaultseppunct}\relax
\EndOfBibitem
\bibitem[Alkauskas and Pasquarello(2011)Alkauskas, and
  Pasquarello]{Pasquarello:09}
Alkauskas,~A.; Pasquarello,~A. \emph{Phys. Rev. B} \textbf{2011}, \emph{84},
  125206\relax
\mciteBstWouldAddEndPuncttrue
\mciteSetBstMidEndSepPunct{\mcitedefaultmidpunct}
{\mcitedefaultendpunct}{\mcitedefaultseppunct}\relax
\EndOfBibitem
\bibitem[Gajdos \latin{et~al.}(2006)Gajdos, Hummer, Kresse, Furthm\"uller, and
  Bechstedt]{Gadjos2006}
Gajdos,~M.; Hummer,~K.; Kresse,~G.; Furthm\"uller,~J.; Bechstedt,~F.
  \emph{Phys. Rev. B} \textbf{2006}, \emph{73}, 045112\relax
\mciteBstWouldAddEndPuncttrue
\mciteSetBstMidEndSepPunct{\mcitedefaultmidpunct}
{\mcitedefaultendpunct}{\mcitedefaultseppunct}\relax
\EndOfBibitem
\bibitem[Onida \latin{et~al.}(2002)Onida, Reining, and Rubio]{Onida}
Onida,~G.; Reining,~L.; Rubio,~A. \emph{Rev. Mod. Phys.} \textbf{2002},
  \emph{601}, 2002\relax
\mciteBstWouldAddEndPuncttrue
\mciteSetBstMidEndSepPunct{\mcitedefaultmidpunct}
{\mcitedefaultendpunct}{\mcitedefaultseppunct}\relax
\EndOfBibitem
\bibitem[Luppi \latin{et~al.}(2016)Luppi, Urdaneta, and Calatayud]{Luppi2016}
Luppi,~E.; Urdaneta,~I.; Calatayud,~M. \emph{J. Phys. Chem. C} \textbf{2016},
  \emph{120}, 5115\relax
\mciteBstWouldAddEndPuncttrue
\mciteSetBstMidEndSepPunct{\mcitedefaultmidpunct}
{\mcitedefaultendpunct}{\mcitedefaultseppunct}\relax
\EndOfBibitem
\bibitem[Casida(1995)]{Casida}
Casida,~M.~E. \emph{Recent Advances in Density Functional Methods}; World
  Scientific, Singapore, edited by D. P. Chong, 1995; Vol.~1\relax
\mciteBstWouldAddEndPuncttrue
\mciteSetBstMidEndSepPunct{\mcitedefaultmidpunct}
{\mcitedefaultendpunct}{\mcitedefaultseppunct}\relax
\EndOfBibitem
\end{mcitethebibliography}

\newpage

\begin{tocentry}
\centering
  \includegraphics[width=0.5\columnwidth,height=4cm,keepaspectratio,clip=]{./homo_ch23-eps-converted-to.pdf}
Highest occupied orbital in SH-(CH$_{2}$)$_{3}$-COOH on the  ($10\bar{1}0$) ZnO  surface.
\end{tocentry}

\end{document}